# Throughput Scaling Of Convolution For Error-Tolerant Multimedia Applications

Mohammad Ashraful Anam[*] and Yiannis Andreopoulos

ABSTRACT

Convolution and cross-correlation are the basis of filtering and pattern or template matching in multimedia signal processing. We propose two throughput scaling options for any one-dimensional convolution kernel in programmable processors by adjusting the *imprecision* (distortion) of computation. Our approach is based on scalar quantization, followed by two forms of tight packing in floating-point (one of which is proposed in this paper) that allow for concurrent calculation of multiple results. We illustrate how our approach can operate as an optional pre- and post-processing layer for off-the-shelf optimized convolution routines. This is useful for multimedia applications that are tolerant to processing imprecision and for cases where the input signals are inherently noisy (error tolerant multimedia applications). Indicative experimental results with a digital music matching system and an MPEG-7 audio descriptor system demonstrate that the proposed approach offers up to 175% increase in processing throughput against optimized (full-precision) convolution with virtually no effect in the accuracy of the results. Based on marginal statistics of the input data, it is also shown how the throughput and distortion can be adjusted per input block of samples under constraints on the signal-to-noise ratio against the full-precision convolution.

EDICS: 2-ALAR Algorithms and Algorithmic Transformations, 2-HDSO Hardware/Software Co-Design and Instruction-Level Architecture Issues

## I. INTRODUCTION

Processing and matching of digital input data in many diverse multimedia applications (e.g. template or signal matching [1] and speaker and music identification [2][3]) is based on digital convolution. Due to its computational complexity, acceleration techniques for convolution and cross-correlation have been studied for a long time. Beyond the tradeoffs between frequency-domain and time-domain convolution [4], research has investigated acceleration of convolution for specific cases. Merhav and Kresch [5] presented a novel method of approximate convolution using discrete cosine transform (DCT) coefficients, which is appropriate only for DCT-domain processing. Di Stefano and Mattoccia [6] presented an accelerated normalized spatial-domain cross-correlation mechanism, with partial check according to an upper bound. Kadyrov and Petrou [7] and Anastasia and Andreopoulos [8] showed that it is possible to perform accelerated 2D integer-to-integer convolution/cross-correlation by piecewise packing of the input image data into floating-point representations. All these cases aim for acceleration of processing by exploiting specific properties of the application, such as transform-domain properties or integer-to-integer processing, and do not explore generic throughput/distortion tradeoffs. Adaptation of throughput according to the output distortion is currently only possible via custom hardware designs [9][10], but all existing approaches provide for *static definition* of the input and kernel precision and no dynamic adjustment of precision can be performed without significant overhead.

[*]Corresponding author. The authors are with the Electronic and Electrical Engineering Department, University College London, Roberts Building, Torrington Place, London, WC1E 7JE, Tel. +44 20 7679 7303, Fax. +44 20 7388 9325 (both authors), Email: uceeman@ee.ucl.ac.uk (M. A. Anam); iandreop@ee.ucl.ac.uk (Y. Andreopoulos). This work was supported by EPSRC, project EP/020015/1 and by a PhD scholarship from the Commonwealth Scholarship Commission.

This paper proposes the formation of two throughput/distortion alternatives as an optional feature for 1D overlap-save convolution realizations in programmable processors. Given the software nature of our approach, the derived designs can be ported to any state-of-the-art multimedia signal processing library [11], any programmable high-end digital signal processor (DSP) [12], and even to graphics processing unit (GPU) realizations [13] used within high-performance multimedia applications. Importantly, unlike related research limited to integer-to-integer processing [7][8] and 2D convolution with $O(N^4)$ complexity [6] (with $N$ the kernel and signal dominant dimension), this is the first work for *non-integer approximate convolution that demonstrates significant acceleration in practice, even when the convolution kernel has $O(N^2)$ or $O(N\log N)$ complexity*, i.e. for 1D time-domain and fast Fourier transform (FFT) based convolution. Finally, the proposed acceleration mechanism is provided externally to the convolution core and it can be selectively disabled if full-precision results are required.

We first review the basics of overlap-save convolution and the proposed approximate computation layer in Section II. The proposed quantization, packing and unpacking method is detailed in Section III. Experimental results validating the obtained throughput scaling within two error-tolerant multimedia applications are provided in Section IV. Concluding remarks are given in Section V.

## II. OVERVIEW OF APPROXIMATE CONVOLUTION REALIZATION

Consider the convolution of two 1D signals[1]:
$$\mathbf{r}_{\text{out}} = \mathbf{s}_{\text{in}} \star \mathbf{k} \Leftrightarrow \forall m: r_{\text{out}}[m] = \sum_{n=0}^{N-1} s_{\text{in}}[m-n]k[n] \qquad (1)$$

The signal with the smallest time-domain support is called *kernel* and the other signal is called *input* [4]. Practical implementations of convolution of an $L$-sample input $\mathbf{s}_{\text{in}}$ with an $N$-sample kernel $\mathbf{k}$ will subdivide the input into $P$ partially-overlapping blocks of $W_{\text{block}}$ samples each (vector $\mathbf{s}$) prior to the actual convolution. Each individual block $\mathbf{s}$ is convolved with the kernel and the resulting blocks ($\mathbf{r}$) are assembled together to give the result of the convolution, $\mathbf{r}_{\text{out}}$. This is the well-known overlap-save method [4], performed for efficient cache utilization and for producing the output in a streaming manner, with latency depending on $W_{\text{block}}$ instead of $L$.

---

[1]Notations: signals are always represented in the time domain by lowercase boldface fonts (indicating vectors) and their elements are shown in brackets, e.g. $\mathbf{s}$ and $s[m]$; $\tilde{\mathbf{s}}, \bar{\mathbf{s}}$ indicate the companded and packed signal $\mathbf{s}$, respectively; $\mathbf{s}_0$ and $\mathbf{s}_1$ are the even and odd samples of $\mathbf{s}$, respectively; $[\![\mathbf{s}]\!], \lfloor \mathbf{s} \rfloor$ and $\lceil \mathbf{s} \rceil$ indicate rounding, floor and ceiling operation respectively (equivalently for scalars); $\|\mathbf{s}\|_\infty$ indicates the $L_\infty$ norm.

As shown in Figure 1, the data-partitioning layer is conventionally positioned at an intermediate layer between the application and the core of the convolution. Our proposal introduces a *middle-layer* (Tier 1.5 in Figure 1) that performs companding and packing/unpacking within each signal block in order to control the processing throughput and the induced distortion of the performed calculation. The advantage of Tier 1.5 is that the system can select an appropriate acceleration mechanism *adaptively*, i.e. based on the input data statistics. At the same time, the convolution core remains untouched; this allows for the usage of any existing high-performance kernel in software or hardware [9]-[13]. We illustrate this aspect by coupling our approach with the state-of-the-art Intel IPP library [11].

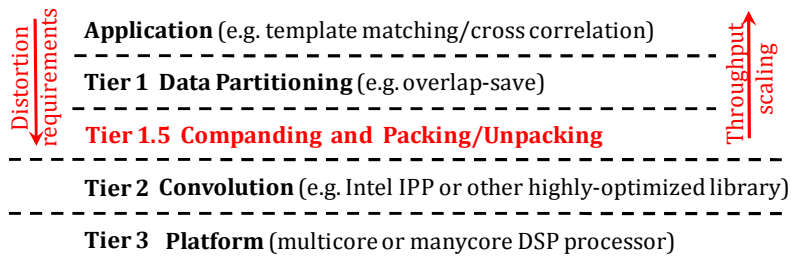

Figure 1. Execution environment of convolution and position of the proposed middle layer.

## III. PROPOSED APPROACH

For presentation simplicity, this section is focusing on odd-length kernels ($N$ is odd); straightforward modifications (or padding the kernel with a single zero) can be applied for even-length kernels. As mentioned, we use blockwise processing of the input signal to allow for memory-efficient streamed processing. In order to ensure the convolution process remains compute-dominated and not memory-dominated, each block is set to produce $W$ results, with $W \geq 2N$ and $W$ even. This leads to each overlapping input block **s** consisting of $W_{\text{block}} = (W + N + 1)$ samples. Importantly, blockwise processing of the input signal $\mathbf{s}_{\text{in}}$ allows for the adaptation of the quantization and packing *during* the execution.

### A. Scalar Quantization and Packing

Initially, both the input signal and the kernel are companded and rounded to integers by:

$$\tilde{\mathbf{s}} = [\![c_s \mathbf{s}]\!], \qquad \tilde{\mathbf{k}} = [\![c_k \mathbf{k}]\!] \tag{2}$$

where $c_s$ and $c_k$ are the chosen companding factors that map the dynamic range of the input signal and kernel to sets $\{-S_q, \ldots, S_q\}$ and $\{-K_q, \ldots, K_q\}$, respectively; for uniform companding, we set $c_s = \frac{S_q}{\|\mathbf{s}\|_\infty}$ and $c_k = \frac{K_q}{\|\mathbf{k}\|_\infty}$. The utilized values for $S_q$ and $K_q$ depend on the error tolerance of each application under consideration; further details on them will be provided in Subsection III.D and in the experimental section. While each of the $P$ signal blocks **s** is companded by (2), the companding of the input kernel can be performed only once at the beginning of the convolution process if $K_q$ remains fixed for the entire process. The maximum possible absolute value of $\tilde{\mathbf{r}} = \tilde{\mathbf{s}} \star \tilde{\mathbf{k}}$ (i.e. the companding range) is

$$R_{\max} = [\![c_s c_k N \|\mathbf{s}\|_\infty \|\mathbf{k}\|_\infty]\!] = [\![N S_q K_q]\!]. \tag{3}$$

The obtained value of $R_{\max}$ is important because it determines the space required in the numerical representation in order to perform error-free packing and unpacking [7][8]. In this work, depending on the throughput and distortion requirements, the selected packing per input block can be: *(i)* a novel *symmetric* packing method proposed in this paper; *(ii)* the *asymmetric* packing method of [7][8]; *(iii)* no packing at all (full-precision overlap-save convolution $\mathbf{r} = \mathbf{s} \star \mathbf{k}$).

In the proposed symmetric packing, the rounded coefficients of each $W_{\text{block}}$-sample input block and the $N$-sample kernel are packed to create two blocks of $\frac{W_{\text{block}}}{2}$ samples and $\lceil \frac{N}{2} \rceil$ samples, respectively by:

$$\bar{\mathbf{s}} = \tilde{\mathbf{s}}_0 + \varepsilon \tilde{\mathbf{s}}_1 \Leftrightarrow \forall \bar{m}: \bar{s}[\bar{m}] = \tilde{s}[2\bar{m}] + \varepsilon \tilde{s}[2\bar{m} + 1] \qquad (4)$$

$$\bar{\mathbf{k}} = \tilde{\mathbf{k}}_0 + \varepsilon^{-1} \tilde{\mathbf{k}}_1 \Leftrightarrow \forall \bar{n}: \bar{k}[\bar{n}] = \tilde{k}[2\bar{n}] + \varepsilon^{-1} \tilde{k}[2\bar{n} + 1] \qquad (5)$$

where: $0 \leq \bar{m} < \frac{W_{\text{block}}}{2}, 0 \leq \bar{n} < \lceil \frac{N}{2} \rceil$ and $\varepsilon$ ($0 < \varepsilon < 1$) is the utilized packing coefficient, which is defined as [7][8]:

$$\varepsilon = (2R_{\max})^{-1} \qquad (6)$$

Notice that this differs from the asymmetric packing method of [7][8], where $M_{\text{asym}}$ parts of the input signal block are packed together ($0 \leq \bar{m} < \lceil W_{\text{block}}/M_{\text{asym}} \rceil$ and, for practical systems [8], $M_{\text{asym}} \in \{2,3\}$):

$$\forall \bar{m}: \bar{s}_{\text{asym}}[\bar{m}] = \sum_{q=0}^{M_{\text{asym}}-1} \varepsilon^q \tilde{s}[\bar{m} + q[W/M_{\text{asym}}]] \qquad (7)$$

and convolution occurs using kernel $\tilde{\mathbf{k}}$ instead of $\bar{\mathbf{k}}$:

$$\bar{\mathbf{r}}_{\text{asym}} = \bar{\mathbf{s}}_{\text{asym}} \star \tilde{\mathbf{k}} \qquad (8)$$

### B. Block Convolution

Once companding and packing has been completed, standard convolution takes place for the $w$th packed signal block $\bar{\mathbf{s}}$, $0 \leq w < P$, by using the packed data to produce the packed output (Tier 2 of Figure 1). Under the symmetric packing, we have ($\forall \bar{m}: W_{\text{start}} \leq \bar{m} < \frac{W_{\text{block}}+N-1}{2}$):

$$\begin{aligned}\bar{\mathbf{r}} &= \bar{\mathbf{s}} \star \bar{\mathbf{k}} \\ &= (\tilde{\mathbf{s}}_0 + \varepsilon \tilde{\mathbf{s}}_1) \star (\tilde{\mathbf{k}}_0 + \varepsilon^{-1} \tilde{\mathbf{k}}_1) \\ &= (\tilde{\mathbf{s}}_0 \star \tilde{\mathbf{k}}_0 + \tilde{\mathbf{s}}_1 \star \tilde{\mathbf{k}}_1) + \varepsilon \times (\tilde{\mathbf{s}}_1 \star \tilde{\mathbf{k}}_0) + \varepsilon^{-1} \times (\tilde{\mathbf{s}}_0 \star \tilde{\mathbf{k}}_1) \\ \Leftrightarrow \bar{r}[\bar{m}] &= \left( \sum_{\bar{n}=0}^{\lfloor N/2 \rfloor} \tilde{s}[2\bar{m} - 2\bar{n}]\tilde{k}[2\bar{n}] + \sum_{\bar{n}=0}^{\lfloor N/2 \rfloor} \tilde{s}[2\bar{m} - 2\bar{n} + 1]\tilde{k}[2\bar{n} + 1] \right) \\ &\quad + \varepsilon \times \sum_{\bar{n}=0}^{\lfloor N/2 \rfloor} \tilde{s}[2\bar{m} - 2\bar{n} + 1]\tilde{k}[2\bar{n}] + \varepsilon^{-1} \times \sum_{\bar{n}=0}^{\lfloor N/2 \rfloor} \tilde{s}[2\bar{m} - 2\bar{n}]\tilde{k}[2\bar{n} + 1]\end{aligned} \qquad (9)$$

with[2]

$$W_{\text{start}} = \begin{cases} 0 & \text{, for } w = 0 \\ \frac{N-1}{2} & \text{, for } w > 0 \end{cases} \qquad (10)$$

Thus, $\bar{\mathbf{r}}$ stemming from convolution with symmetric packing contains three *entangled* outputs, the *base* output – multiplied by $\varepsilon^0 = 1$, and two *side* outputs – multiplied by $\varepsilon$ (side-$\varepsilon$) and $\varepsilon^{-1}$ (side-$\varepsilon^{-1}$). We can derive two convolution results via these three outputs as detailed in the following subsection.

### C. *Unpacking and Inverse Companding*

The results of the $w$th block $\bar{\mathbf{s}}$, $0 \leq w < P$, are unpacked, inverse companded, and combined to form the final results.

---

[2] As in overlap-save convolution, where the first $N - 1$ outputs are discarded beyond the first window processing [4], during processing and unpacking we discard the first $W_{\text{start}}$ outputs since they are not used.

Unlike asymmetric packing [i.e. (8)] where the convolution outputs $\bar{\mathbf{r}}_{\text{asym}}$ can be unpacked following $M_{\text{asym}}$ successive iterations [(5) and (6), [8]], under convolution with symmetric packing the three terms are unpacked by the following set of equations, which are repeated for each $\bar{m}$, $W_{\text{start}} \leq \bar{m} < \frac{W_{\text{block}}}{2}$, with $W_{\text{start}}$ defined by (10).

Initially, each sample is brought to the positive domain of floating point by:

$$u[\bar{m}] = \bar{r}[\bar{m}] + R_{\text{safe}} \qquad (11)$$

where $R_{\text{safe}} \triangleq R_{\max}(\varepsilon + 1 + \varepsilon^{-1}) + u_{\text{sys}}\varepsilon^{-1}$ [with $R_{\max}$ derived by (3)] and $u_{\text{sys}}$ a system-dependent parameter, which is derived in the experimental section. Then the terms entangled by $\varepsilon$ and by $\varepsilon^{-1}$ are extracted by:

$$r_{\text{side}-\varepsilon} = \begin{cases} R_{\max}, & \text{for } \bar{m} = W_{\text{start}} \\ \varepsilon^{-1}(u[\bar{m}-1] - \lfloor u[\bar{m}-1] \rfloor), & \text{for } \bar{m} > W_{\text{start}} \end{cases} \qquad (12)$$

$$r_{\text{side}-\varepsilon^{-1}} = \lfloor \varepsilon u[\bar{m}] \rfloor \qquad (13)$$

These results are used in order to produce one of the outputs by:

$$\tilde{r}[2\bar{m}] = \lfloor r_{\text{side}-\varepsilon^{-1}} + r_{\text{side}-\varepsilon} - 2R_{\max} \rfloor \qquad (14)$$

The term $-2R_{\max}$ in (14) brings the output $\tilde{r}[2\bar{m}]$ to its original range, since both $r_{\text{side}-\varepsilon}$ and $r_{\text{side}-\varepsilon^{-1}}$ were brought to the positive domain by adding $R_{\text{safe}}$. Finally, the other output of the convolution is derived by extracting the component multiplied by $\varepsilon^0$ and again subtracting $R_{\max}$ to bring the value to its original range:

$$\tilde{r}[2\bar{m}+1] = \lfloor \lfloor u[\bar{m}] \rfloor - (\varepsilon^{-1} r_{\text{side}-\varepsilon^{-1}}) - R_{\max} \rfloor \qquad (15)$$

Inverse companding is applied to recover the final results:

$$\mathbf{r} = (c_s c_k)^{-1} \tilde{\mathbf{r}} \qquad (16)$$

The output from the first block ($w = 0$) consists of $(W + N - 1)$ samples and is placed directly in the output ($r_{\text{out}}[m] = r[m]$). For $w > 0$, $W$ output samples are produced, which are placed in the output by ($\forall m: 0 \leq m < W$):

$$r_{\text{out}}[N - 1 + W \times w + m] = r[m + 2W_{\text{start}} + 2] \qquad (17)$$

## D. Discussion

Table 1 summarizes the features of the proposed companding and packing (symmetric and asymmetric) in comparison to conventional processing. The arithmetic and memory requirements are calculated assuming the minimum block size: $W = 2N \Rightarrow W_{\text{block}} = 3N + 1$; larger input block sizes will have proportionally-higher requirements for all methods. The floating-point operations (FLOP) under frequency-domain convolution stem from the FFT FLOP approximation of Franchetti *et al* [13]. The arithmetic and memory cost of companding, packing and unpacking is included by counting their additions and multiplications; floor operations are performed by data conversion and have no arithmetic cost. The derivation of these formulas is trivial and is omitted for brevity of description. The table shows that symmetric packing is expected to be the fastest method under both time-domain and frequency-domain convolution. These high-level estimates will be coupled with experiments in order to assess the throughput gains obtained in practice.

***Companding ranges*:** The proposed symmetric packing utilizes three positions within the numerical representation [at $\varepsilon^0$, $\varepsilon^1$ and $\varepsilon^{-1}$, as shown by (9)]. On the other hand, the asymmetric packing of (7) utilizes $M_{\text{asym}}$ positions, at $\varepsilon^q$ $\forall q \in \{0, \ldots, M_{\text{asym}} - 1\}$ [7][8]. Hence, from tight-packing theory (Proposition 1 of [8]), for error-free unpacking we get upper bounds (under double-precision floating point representation) shown in the "Companding range" row of Table 1.

***Distortion:*** The source of imprecision in the proposed approach is the quantization of (2). Assuming the input signal and kernel are modeled by zero-mean, independent, identically distributed (iid), random variables with standard deviations $\sigma_s$ and $\sigma_k$ (respectively), and the companding and rounding noise for the signal and kernel is zero-mean iid with standard deviation $\sigma_{v_s}$ and $\sigma_{v_k}$, the expected noise power stemming from companding and rounding is:

$$D_r = N\left[(\sigma_s \sigma_{v_k})^2 + (\sigma_k \sigma_{v_s})^2 + (\sigma_{v_k} \sigma_{v_s})^2\right] \tag{18}$$

This is derived for each output $r[m]$ by expressing the impact of the companding and rounding process of (2) in affine form that contains all the noise sources [14] and subtracting from the equivalent expression for the noiseless (full-precision) operation of (1). We can convert $D_r$ into signal-to-noise ratio (SNR) against the full-precision result by:

$$\text{SNR}_r = 10 \log_{10} \frac{N(\sigma_s \sigma_k)^2}{D_r} \tag{19}$$

Assuming (2) introduces uniform quantization noise, we have: $\sigma_{v_s} = \frac{1}{c_s \sqrt{12}}$ and $\sigma_{v_k} = \frac{1}{c_k \sqrt{12}}$ with $c_s = \frac{S_q}{\|\mathbf{s}\|_\infty}$ and $c_k = \frac{K_q}{\|\mathbf{k}\|_\infty}$. This leads to the SNR expression of Table 1, which is the expected SNR if the inputs are iid with known or estimated standard deviations $\sigma_s$ and $\sigma_k$. As we shall show in the experimental section, the SNR estimate given in Table 1 can be used even under non-iid conditions to predict the *SNR increase* when increasing the companding range ($S_q$ and $K_q$) if we use $\sigma_s$ and $\sigma_k$ derived from the marginal distribution of the input sources. This provides an analytic mechanism for controlling the distortion according to the error tolerance of the convolution of each input block.

Table 1. Comparison of different approaches for input signal block $W_{\text{block}} = 3N + 1$ and companding ranges $S_q$, $K_q$.

| Method | Conventional (full precision) convolution | Companding and symmetric packing | Companding and asymmetric packing [7][8] |
|---|---|---|---|
| Floating-point operations (FLOP = additions, multiplications) | Time: $4N^2$<br><br>Frequency: $(45N+15)\log_2(3N+1)+3N+1$ | Time: $N^2 + N$<br><br>Frequency: $(22.5N+7.5)\log_2(3N+1)+5N+1$ | $M_{\text{asym}} = 2$:<br>Time: $2.5N^2 - N$<br>Frequency: $(30N+15)\log_2(2N+1) + \frac{39}{2}N + \frac{13}{2}$<br><br>$M_{\text{asym}} = 3$:<br>Time: $2N^2 - \frac{2}{3}N$<br>Frequency: $(25N+15)\log_2(\frac{5}{3}N+1) + \frac{62}{3}N + 7$ |
| Memory (samples) | Time: $4N + 1$<br>Freq.: $6N + 2$ | Time: $2N + 1$<br>Freq.: $3N + 1$ | $M_{\text{asym}} = 2$: Time: $\lceil \frac{5}{2}N \rceil$, Freq.: $3N + 1$<br>$M_{\text{asym}} = 3$: Time: $2N + 1$, Freq.: $2N + 1$ |
| Companding range $R_{\max}$ | Not applicable | $[\![NS_q K_q]\!] \leq 97667$ | $M_{\text{asym}} = 2$: $[\![NS_q K_q]\!] \leq 43165096$<br>$M_{\text{asym}} = 3$: $[\![NS_q K_q]\!] \leq 97667$ |
| SNR under iid inputs with std $\sigma_s$ & $\sigma_k$ and iid quantization noise with std $\frac{\|\mathbf{s}\|_\infty}{S_q\sqrt{12}}$ & $\frac{\|\mathbf{k}\|_\infty}{K_q\sqrt{12}}$ | Not applicable | $10 \log_{10} \dfrac{144(\sigma_s \sigma_k S_q K_q)^2}{12\left[(\sigma_s K_q \|\mathbf{s}\|_\infty)^2 + (\sigma_k S_q \|\mathbf{k}\|_\infty)^2\right] + (\|\mathbf{s}\|_\infty \|\mathbf{k}\|_\infty)^2}$ | |

## IV. EXPERIMENTAL RESULTS

We implemented the proposed approach in an Intel i5 540M 2.5GHz processor (Windows 7 64bit SP1, single threaded, Intel C/C++ compiler version 12, switches: `/Ox /Ot /QxHost`). We selected $u_{sys} = 3.1193 \times 10^{-11}$ based on the machine precision measurement of [7]. The following process is performed for companding and packing:

*Initialization:* We find $\|\mathbf{k}\|_\infty$ and decide on the companding of the input and the kernel [via (2)] by defining $S_q$ and $K_q$ either based on distortion requirements or based on application requirements (see Subsections IV.B-IV.D for examples). The kernel is companded uniformly via (2) using $c_k = \frac{K_q}{\|\mathbf{k}\|_\infty}$ and packed using (5). The following steps are then repeated for every input block **s**.

- *Block companding:* We find $\|\mathbf{s}\|_\infty$ and compand the block via (2) using $c_s = \frac{S_q}{\|\mathbf{s}\|_\infty}$.
- *Block packing:* We set $\varepsilon$ via (6) with $R_{max}$ set via (3). Packing of the input block is performed by (4) for symmetric packing and by (7) for asymmetric packing with $M_{asym} \in \{2,3\}$.

Even though uniform companding is performed per block, the above process adapts the companding parameters $c_k$ and $c_s$ according to kernel and each input block's maximum value. Importantly, the implementation can straightforwardly switch between asymmetric, symmetric and no packing with no overhead, in order to provide adaptive convolution for error-tolerant applications [16], while the convolution core remains unaffected.

### A. Throughput Tests using the Intel Integrated Performance Primitives (IPP) 7.0

To examine the acceleration offered by the proposed approach, we created test input signals of $L = 8192 \times 2^{10}$ samples to be convolved with filter kernels of $N = 800$ samples. Concerning block processing, we varied the block size between $W \in \{2,4,8,16,32,64\} \times 2^{10}$ samples. The input signal and kernel values were randomly generated between $[-128.0, 128.0]$ with double-precision floating-point accuracy. We set $S_q = K_q = 16$ for symmetric packing and asymmetric packing with $M_{asym} = 3$; furthermore, we set $S_q = K_q = 256$ for asymmetric packing with $M_{asym} = 2$. The Intel IPP [11] (routine `ippsConv_64f()`, single-threaded) was used for all Tier-2 convolution.

The results are given in Figure 2 in terms of Msamples/s ($10^6$ output samples per second) achieved by each case. A summary of the obtained throughput increase and the resulting distortion against the full-precision calculation is given in Table 2. Even though the performance gains vary according to the size of the input block, the proposed approach provides significant throughput increase even for the case of $O(N\log N)$ complexity. Using the distortion estimation of Table 1, we find that companding and symmetric packing or asymmetric packing with $M_{asym} = 3$ achieves $\text{SNR}_r = 27.1\text{dB}$ against the full-precision (conventional) calculation; companding and asymmetric packing with $M_{asym} = 2$ achieves $\text{SNR}_r = 51.2\text{dB}$. This is in good agreement with the average SNR measured from the experiments (Table 2).

Finally, to examine the impact of the proposed approach when using convolution kernels of different size, we varied the convolution kernel length, $N$, under input block $W = 32 \times 2^{10}$ samples, and measured the obtained throughput for each case. The results are presented in Figure 3; Figure 3(b) shows the theoretical throughput obtained by dividing the total output samples with the FLOP per case (from Table 1) followed by linear scaling to match the range of Figure 3(a). Evidently, symmetric packing achieves significant acceleration in comparison to full-precision convolution and to asymmetric packing. This agrees with the theoretically-predicted performance ranking based on FLOP counts.

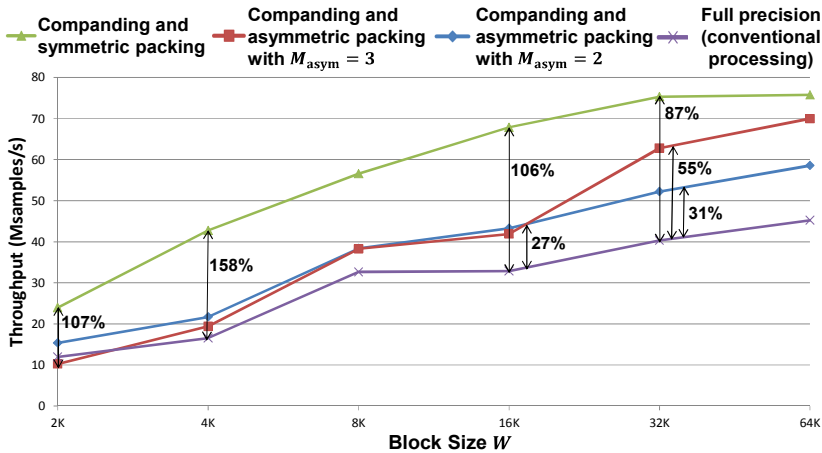

Figure 2. Throughput of convolution under different packing methods (higher is better) with the underlying convolution realization provided by Intel IPP routine `ippsConv_64f()`[11] for all approaches.

Table 2. Minimum and maximum percentile throughput increase and average SNR measured for each method of Figure 2 against the conventional (full-precision) IPP realization.

| Packing Method | Min. (%) | Max. (%) | SNR (dB) |
|---|---|---|---|
| Asymmetric $M_{asym} = 2$ | 18 | 32 | 51.3 |
| Asymmetric $M_{asym} = 3$ | -14 | 55 | 27.5 |
| Symmetric | 52 | 158 | 27.5 |

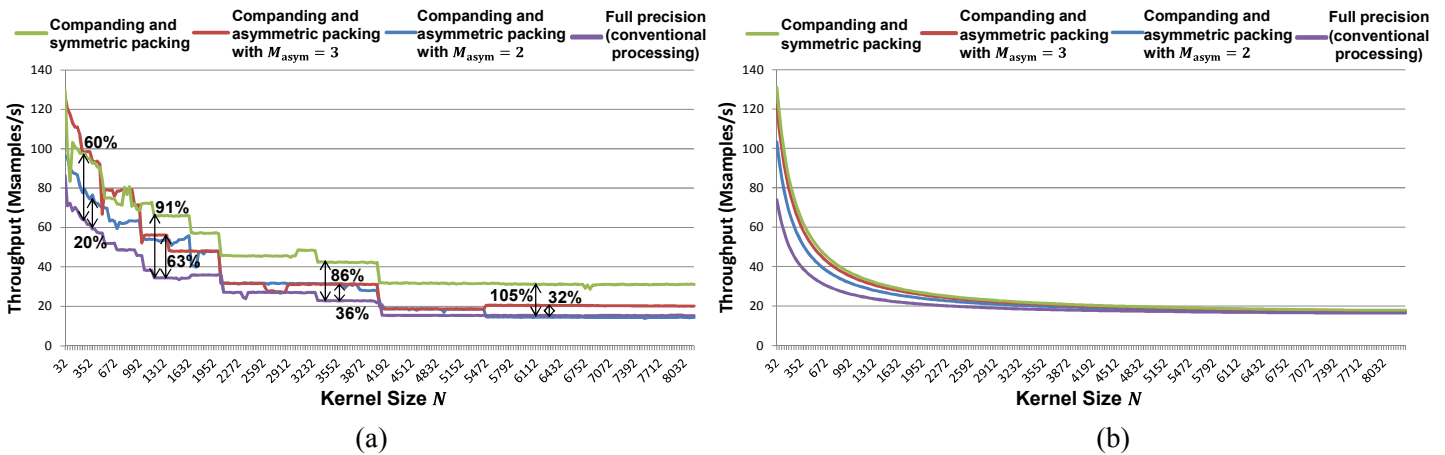

(a)          (b)

Figure 3. Throughput of each method (higher is better) under varying filter kernel size, from $N = 32$ to $N = 8192$ samples; (a) experiments with Intel IPP (the variability is due to IPP convolution core's internal reconfiguration); (b) theoretical calculation of throughout based on the FLOP counts of Table 1.

### B. *Error-tolerant Application 1: Music Matching System*

A recently-proposed music matching system that identifies cover songs [3] was selected as the first test case. For each input song to be identified, the system works by extracting beat and tempo data and then matching it to a pre-calculated beat and tempo database via cross correlation. Matlab code for this and the sample data were collected from the authors' site [3]. Given that this implementation is dominated by the cross-correlation operations [3], the only modification performed was the replacement of the Matlab `xcorr()` function call with our convolution implementation. Thus, in this case each input block of the convolution corresponds to a song's beat and tempo data and each convolution kernel is the beat and tempo data of a song of the database. The settings used for our experiments were: average beat rate 120 beats-per-minute, chroma element central frequency 200Hz [3]. Concerning our implementation, we set $S_q = K_q = 8$. Table 3 demonstrates that these settings yielded the same matching accuracy for all methods (53.75% match), while providing up to 112% increase in throughput in comparison to the full-precision (conventional) Intel IPP implementation, which in turn accelerates the processing by an order of magnitude in comparison to the Matlab implementation of the authors [3]. Interestingly, asymmetric packing with $M_{asym} = 3$ turns out to be marginally slower than with $M_{asym} = 2$ because, for small kernel sizes, the cost of the additional packing and unpacking may overturn the gain in throughput, as shown also by Figure 2.

Table 3. Matching accuracy vs. cross-correlation throughput for the music matching system of Ellis *et al* [3].

| Method | Matching Accuracy | Throughput (MSamples/s) |
|---|---|---|
| Full-precision Intel IPP | 53.75% | 6.54 |
| Companding and asymmetric packing, $M_{asym} = 2$ | 53.75% | 13.71 |
| Companding and asymmetric packing $M_{asym} = 3$ | 53.75% | 13.49 |
| Companding and symmetric packing | 53.75% | 13.90 |

*C. Error-tolerant Application 2: MPEG-7 Audio Descriptor Calculation*

MPEG-7 [15] is an ISO/IEC standard providing a rich set of standardized tools to describe multimedia content, thus enabling effective and efficient access (search, browsing and filtering) to media streams. For audio stream processing, one of the initial parts of the MPEG-7 processing workflow is determining if a multi-channel audio is actually mono. Furthermore, in order to synchronize multi-channel audio, it is imperative to calculate the relative delay between multiple channels. Signal blocks from each audio channel are cross-correlated to perform these tasks. To investigate this application, Matlab code for MPEG-7 compliant processing of audio sources (including example audio clips) were collected from an MPEG-7 multimedia software resources site [15]. The standard Matlab cross correlation `xcorr()` was replaced with our packed implementation of cross-correlation to calculate the throughput increase and the incurred loss in accuracy in this application. The default sample block size of 50ms was used for the first test and then varied to 100ms and 200ms to verify the results for other block sizes. We set $S_q = K_q = 32$ for this case.

Table 4. Results' accuracy vs. cross-correlation throughput for the MPEG7 audio descriptor calculation [15].

| Method | Block: 100ms (4410 samples) | | | |
|---|---|---|---|---|
| | Relative Delay Error | Cross-channel Correlat. Error | Mono/Stereo Correct Detection | MSamples/s |
| Full-precision Intel IPP | 0.00% | 0.00% | 100% | 5.68 |
| Companding and asymmetric packing, $M_{asym} = 2$ | 0.00% | 0.03% | 100% | 5.48 |
| Companding and asymmetric packing, $M_{asym} = 3$ | 0.00% | 0.03% | 100% | 5.49 |
| Companding and symmetric packing | 0.00% | 0.03% | 100% | 15.57 |

Indicative results from the sample "DemoBad.wav" are shown in Table 4 (other audio files and block sizes yielded similar results). It is evident that our packed implementation provides considerable throughput increase over conventional Intel IPP (175%) while suffering little or no loss of accuracy (no error for relative delay and mono/stereo audio detection and only 0.03% error for cross-channel correlation). Interestingly, asymmetric packing suffers at this signal and filter sizes because the packing/unpacking overhead negates the gain achieved by packing only the signal.

### D. Throughput Scaling under a Lower Limit on the SNR per Input Block

We examine the possibility of dynamically scaling the processing throughput under a lower limit on the required SNR of the convolution of each input block. This requires the analytic estimation of the SNR obtained under different companding parameters $S_q$ and $K_q$, which is illustrated here for the two applications of the previous two subsections.

The samples of the input features for the music matching (or the input audio – for the MPEG-7 descriptor calculation) have significant statistical dependencies due to their temporal correlation. Thus, using affine forms for estimating the mean and variance of $r[m]$ and $\tilde{r}[m]$ [as performed for the derivation of (18)] can quickly lead to intractable analytic expressions due to the cross terms appearing between the moments of random variables modeling temporally adjacent samples. Hence, we opt to directly model the *marginal* cumulative distribution function (CDF) of temporally adjacent input samples, with the square root of its second moment giving the value for $\sigma_s$ and $\sigma_k$ to be used in $D_r$ of (18). This will derive the error power and SNR under marginal CDF modeling, which will be underestimating the experimentally-observed SNR. However, this underestimation is amortized when assessing the SNR *difference* under varying $S_q$ and $K_q$; alternatively, it can be calibrated via a single SNR measurement, as shown in the following.

For the two applications demonstrated earlier, we illustrate the modeling of the marginal statistics of representative inputs in Figure 4. The experimental cumulative histograms represent the cumulative frequencies of several representative inputs within several 16-sample sets. We experimented with several CDFs and the best result after fitting with the expectation maximization algorithm is reported for each case. Our choices for the marginal CDFs, i.e. bivariate Weibull and Laplacian, agree with past results on the modeling of audio features [18] and audio samples [19].

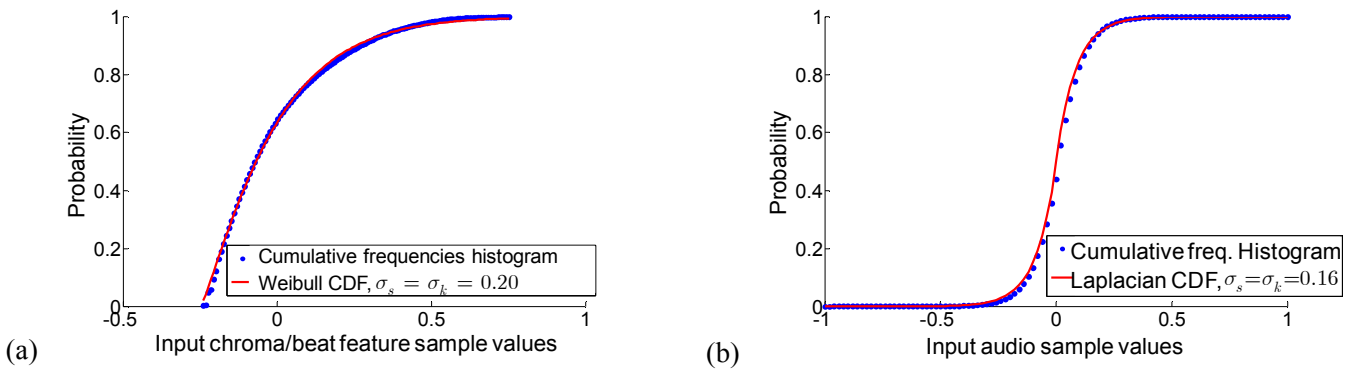

Figure 4. CDF fitting to the histogram of cumulative frequencies of several temporally adjacent sets of inputs; (a) Weibull CDF ($\lambda = 0.248$ and $\kappa = 1.191$) for the music matching application of Subsection IV.B; (b) Zero-mean Laplacian CDF for the MPEG-7 feature processing application of Subsection IV.C.

Using the obtained $\sigma_s$ and $\sigma_k$ values from the fitting, the results of the SNR expression of Table 1 under varying $S_q$ and $K_q$ (and for several inputs) are found in Figure 5. All model SNR results are calibrated via a single measurement at the coarsest quantization ($S_q = K_q = 8$) in order to provide an absolute SNR estimator per application. The figure also indicates the regions where different companding and packing options can be achieved. The results show that the SNR formula of Table 1 can be used to predict the distortion when switching from one companding option to another.

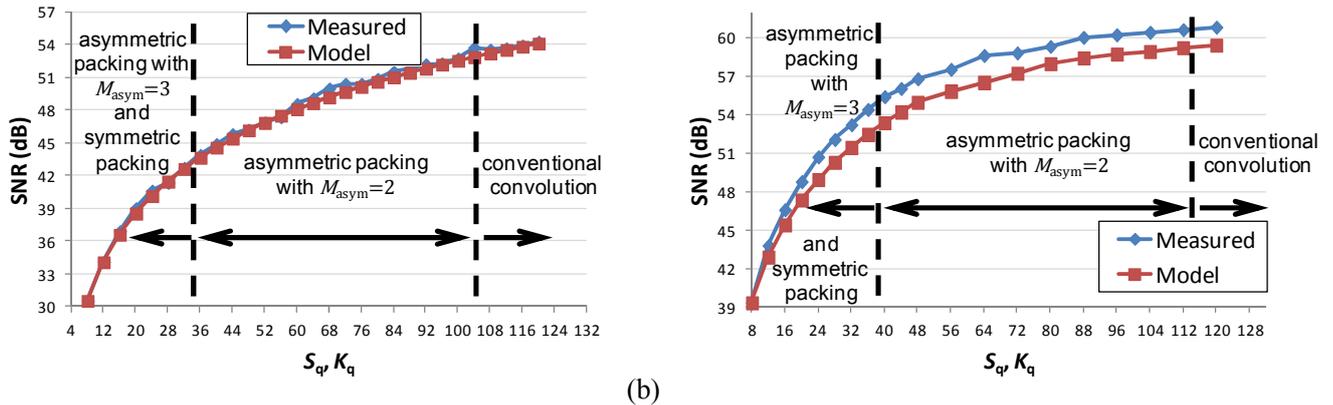

Figure 5. Average SNR for: (a) the music matching application (beat cross-correlation); (b) the MPEG-7 delay calculation via cross-channel correlation. The model results are calibrated via one SNR measurement at $S_q = K_q = 8$.

We illustrate the applicability of such a scheme in the dynamic adaptation of the convolution throughput for the music matching application. Figure 6(a) demonstrates an example of varying the minimum SNR requirement for each cross-correlation of the input beat and chroma data of a song with the beat and tempo database; hence, "Block Number" indicates beat & chroma entry of each song in the database (only 9 are shown). Figure 6(b) shows the throughput obtained by symmetric, asymmetric packing ($M_{\text{asym}} = 2$) or full-precision processing. This adjustment is done by using the model of Figure 5(a) to first establish which packing (if any) allows for SNR surpassing the requirement and then selecting the *maximum* $S_q$ and $K_q$ that allow for this packing. If no companding and packing approach allows for the minimum SNR requirement to be met, then full-precision convolution is performed (i.e. SNR is infinity). Because only three options are available per block, the obtained SNR can be significantly higher than the minimum requested.

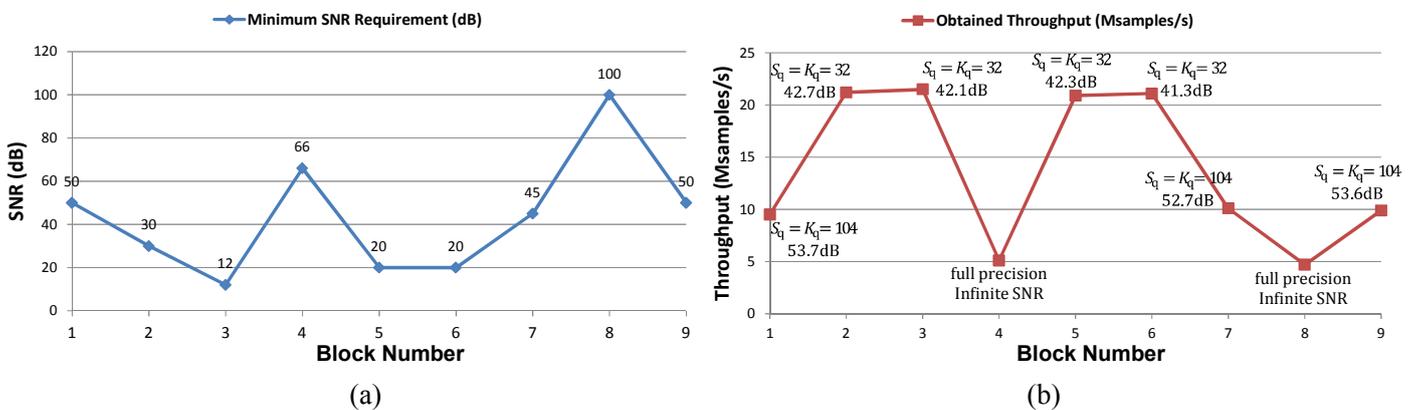

Figure 6. Companding and packing adaptation according to SNR constraints per convolution block.

## V. Conclusion

We propose (and make available online [16]) an operational approach that scales the throughput of generic convolution and cross-correlation by increasing the imprecision (distortion) of the results. This can be used in multimedia applications where higher throughput is desired and certain imprecision can be tolerated (or is inherently present) due to noise in the input data (error tolerant multimedia processing). The possibility of selecting amongst three alternatives per convolution window can provide for software realizations that balance between throughput, memory and accuracy requirements, as detailed in the summary of Table 1. The proposed method can be applied as an optional layer for any high-performance convolution kernel as it operates externally to the kernel code. We validated this approach with the state-of-the-art Intel IPP convolution kernel in a digital music matching application and in MPEG-7 descriptor computations, where we demonstrated significant processing-throughput gain with only marginal loss of accuracy.